\documentstyle[aps,preprint,tighten,floats,epsfig,amsfonts]{revtex}
\begin{document}
\newcommand{\nc}{\newcommand}
\nc{\be}{\begin{equation}}
\nc{\ee}{\end{equation}}
\nc{\bea}{\begin{eqnarray}}
\nc{\eea}{\end{eqnarray}}
\nc{\bib}{\bibitem}
\nc{\1}{16^3\times{32}}
\nc{\2}{24^3\times{36}}
\nc{\al}{\alpha}
\nc{\bt}{\beta}
\nc{\sutwo}{SU_{c}(2)}
\nc{\suthree}{SU(3)}
\nc{\sucthree}{SU_{c}(3)}
\nc{\sun}{SU(N)}
\nc{\sucn}{SU_{c}(N)}
\nc{\xht}{\hat{x}}
\nc{\yht}{\hat{y}}
\nc{\zht}{\hat{z}}
\nc{\tht}{\hat{t}}
\nc{\n}{(n+1)}
\nc{\m}{(m+1)}
\nc{\bpt}{(x,y,z,t)}
\nc{\bptpone}{(x,y,z+1,t+1)}
\nc{\bptptwo}{(x,y,z+2,t+2)}
\nc{\bptpthree}{(x,y,z+3,t+3)}
\nc{\bptplast}{(x,y,z+m,t+m)}
\nc{\obyo}{1\times{1}}
\nc{\mbym}{m\times{m}}

\draft
\preprint{\vbox{\hfill\rm ADP-00-04/T392
}}

\title{\bf General Algorithm For Improved Lattice Actions on Parallel
Computing Architectures}

\author{Fr\'{e}d\'{e}ric D.R. Bonnet\footnote{
E-mail:~fbonnet@physics.adelaide.edu.au ~$\bullet$~ Tel:
+61~8~8303--3428 ~$\bullet$~ Fax: +61~8~8303--3551}$^{1}$,
Derek B. Leinweber\footnote{E-mail:~dleinweb@physics.adelaide.edu.au
~$\bullet$~ Tel:
+61~8~8303--3423 ~$\bullet$~ Fax: +61~8~8303--3551 \hfill\break
\null\quad\quad 
WWW:~http://www.physics.adelaide.edu.au/theory/staff/leinweber/}$^{1}$,
and
Anthony G. Williams\footnote{E-mail:~awilliam@physics.adelaide.edu.au 
~$\bullet$~ Tel:
+61~8~8303--3546 ~$\bullet$~ Fax: +61~8~8303--3551 \hfill\break
\null\quad\quad
WWW:~http://www.physics.adelaide.edu.au/cssm/}$^{1,2}$.}
\address{ $^1$ Special Research Center for the Subatomic Structure of Matter 
(CSSM) and Department of Physics and Mathematical Physics, University of 
Adelaide 5005, Australia.}
\address{ $^2$ Department of Physics and SCRI, Florida State University, 
Tallahasse, FL 32306.}
\date{\today}
\maketitle

\begin{abstract}

Quantum field theories underlie all of our understanding of the
fundamental forces of nature.  The are relatively few first principles
approaches to the study of quantum field theories [such as quantum
chromodynamics (QCD) relevant to the strong interaction] away from the
perturbative (i.e., weak-coupling) regime.  Currently the most common
method is the use of Monte Carlo methods on a hypercubic space-time
lattice.  These methods consume enormous computing power for large
lattices and it is essential that increasingly efficient algorithms be
developed to perform standard tasks in these lattice calculations.
Here we present a general algorithm for QCD that allows one to put any
planar improved gluonic lattice action onto a parallel computing
architecture.  High performance masks for specific actions (including
non-planar actions) are also presented.  These algorithms have been
successfully employed by us in a variety of lattice QCD calculations
using improved lattice actions on a 128 node Thinking Machines CM-5.

{\underline{Keywords}}: quantum field theory; quantum chromodynamics;
improved actions; parallel computing algorithms.

\end{abstract}

\newpage
\section{Introduction}

It is almost universally accepted that Quantum Chromodynamics (QCD) is the
underlying quantum field theory of the strong interaction
\cite{Peskin:1995,Muta:1987}, which binds
atomic nuclei and fuels the sun and the stars.  Strongly interacting
particles are referred to as hadrons, which include for example protons
and neutrons that make up atomic nuclei as well as a wide variety
of particles that are produced in particle accelerators
and from astrophysical sources.  These hadrons are made up of quarks
and gluons, which are the underlying constituents in QCD.  The quarks
are spin-1/2 particles (i.e., fermions) and the gluons are massless
spin-1 particles (i.e., gauge bosons).  The quarks interact strongly
through their ``colour'' charge through the exchange of gluons.
The 8 gluons of \suthree\ (i.e., one for each generator of \suthree)
themselves carry colour and hence interact with themselves
as well as with the quarks.  This is the essential difference between
QCD and the corresponding theory of photons and electrons
referred to as quantum electrodynamics (QED) and has far reaching
consequences since the theories have entirely different behavior.

The are very few first-principles methods for studying QCD in the
nonperturbative low-energy regime. The most widely used of these
is the so-called Lagrangian-based lattice field theory, which formulates
the field theory on a space-time lattice \cite{Rothe:1992,Montvay:1994}.
An alternative lattice approach is based on the Hamiltonian formulation
of quantum field theory and makes use of cluster decompositions and again
Monte Carlo methods to carry out the simulations\cite{Schutte:1997du}.
In addition, there are numerous studies based on a light-front
formulation of QCD \cite{Brodsky_etal} and much use has been
made of Schwinger-Dyson equations\cite{DSE_review}
to assist with the construction of QCD-based quark models.

The Lagrangian-based lattice technique simulates the functional
integral using a four-dimensional hypercubic Euclidean spacetime
lattice together with Monte Carlo methods for generating an ensemble
of gluon field configurations with the appropriate Boltzmann
distribution $\exp(-S_G)$, where $S_G$ is a discretized form of the
QCD gluon action on the hypercubic lattice.  The simplest
discretizations of the QCD action involve only nearest neighbours on
the lattice and have ${\cal O}(a^2)$ errors, where $a$ is the lattice
spacing.  Improved actions represent a major advance for the field of
lattice gauge theory, where by using increasingly non-local
discretizations of the QCD action we can obtain the same accuracy with
far fewer lattice points and hence far less computational time and
effort.  The purpose of the present work is to describe an algorithm
which allows us to implement an arbitrarily improved (i.e.,
arbitrarily non-local) action in an efficient way.  For further details
on the state of the art lattice QCD techniques see for example
Ref.~\cite{proc:latt97}.  Another related and equally important
advance is the technique of nonperturbative improvement (e.g.,
mean-field improvement) which corrects for some of the major
nonperturbative effects (the so-called tadpole contributions) and
hence more quickly brings the lattice results to their continuum form
by improving the matching with perturbation theory at a given lattice
spacing $a$~\cite{mackenzie}.  It is the combination of improved
actions and nonperturbative improvement that together have come to
represent a significant advance for the
field~\cite{proc:latt97,lepage}.

Lattice QCD is based on a Monte Carlo treatment of the path integral
formulation, which makes it a computationally demanding method for
calculating physical observables.  The gluon field is represented by
$3\times 3$ complex $SU(3)$ matrices, where there is one such $SU(3)$
matrix associated with every link on the lattice.  The links lie only
along one of the four Cartesian directions and join neighbouring
lattice sites.  Since all lattice links require identical numerical
calculations, lattice gauge theory is ideally suited for parallel
computers.

\begin{figure}[hbt]
\centering{\
        \epsfig{angle=0,figure=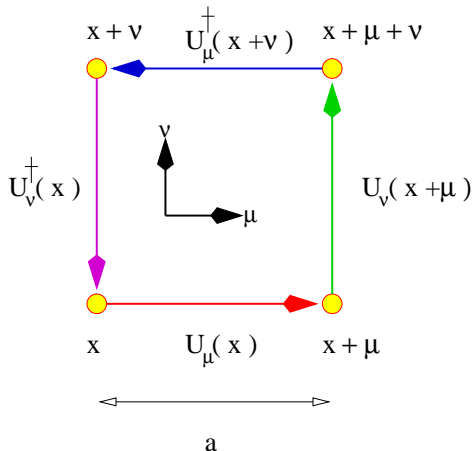,height=6cm} }
\parbox{130mm}{\caption{The $1\times 1$ plaquette $U_{\rm{sq}}(x)$ 
with base at $x$ lying in the $\mu\nu$-plane.  The lattice spacing 
is denoted by $a$.}
\label{plaquette}}
\end{figure}

There are various types of improved actions and, as explained above,
these are all based on the idea of eliminating the discretization
errors that occur when passing from continuum physics to the
discretized lattice version.  The simplest (i.e., non-improved) gluon
action is the so called {\em standard Wilson} action and consists of
$\obyo$ Wilson loops or as they are frequently called plaquettes.  We
shall often refer to the Wilson loops used to build up lattice actions
as plaquettes.  The need to build the gluon action out of closed loops
arises from the need to maintain exact $SU(3)$ gauge invariance in the
discrete lattice action. This $\obyo$ loop action was first proposed
by Wilson~\cite{wilson} in the early 70's and has been used
extensively over the years.  It consists of taking an arbitrary
starting site, say $x$, on the lattice and stepping around a $\obyo$
loop until returning to the starting point $x$.  The $\obyo$ Wilson
loop is illustrated in Fig.~\ref{plaquette}.

Improving the standard Wilson action is achieved by making use
of larger loops (e.g., $1\times 2$, $2\times 2$, etc.)
in the lattice gluon action\cite{Fiebig} to eliminate finite lattice
spacing artifacts to a given order in $a^2$~.  For an elegant and
detailed discussion of these topics see Ref.~\cite{lepage}.

In this article we present an efficient and completely general algorithm
that permits 
one to calculate any improved planar lattice action at any desired level
of improvement.  By ``planar'' here we mean that we will consider
actions containing two-dimensional loops of arbitrary size which
lie in any of the Cartesian planes, (i.e., the $x-y$, $x-z$, $x-t$,
$y-z$, $y-t$, or $z-t$ plane).   This algorithm has been used in
a wide variety of improved action lattice simulations to date
\cite{proc:latt97}.
For example, it has been used in studies of the topological structure
of the QCD vacuum and the calibration of the various cooling and smearing
techniques~\cite{bonnet}, the study of
discretization errors in the Landau gauge on the lattice~\cite{bowman},
and studies of the static quark potential~\cite{zanotti}.
It is currently being used in studies of the gluon propagator~\cite{bowman2}
and highly improved actions~\cite{bilson-thompson}. 

In Sec.~\ref{action}, we briefly describe the tree--level improved
action that we have been using in our
calculations~\cite{bonnet,bowman,zanotti,bowman2,bilson-thompson} with
our algorithm on a Thinking Machines CM-5.  Sec.~\ref{masking_Wilson}
gives two possible ways of using the technique for the standard Wilson
action.  The form of the algorithm appropriate for the first level of
improvement (i.e., involving a combination of the elementary square
$1\times 1$ plaquette and the rectangular $1\times 2$ plaquette) is
given in Sec.~\ref{masking_improved}.  Then in
Sec.~\ref{maskingsupimpsu3} we present the general algorithm suitable
for an arbitrarily non-local action, (i.e., for an $n\times m$ Wilson
loop with $n$ and $m$ being arbitrary positive integers).
Sec.~\ref{nonplanar} addresses non-planar issues encountered in
nesting specific planar actions.  Actions involving non-planar loops
are also addressed.  Finally, in Sec.~\ref{Conclusion} we present our
summary and conclusions.

\section{Gauge Action, Masking, and Parallel Computing}
\label{action}

\subsection{Lattice Gauge Action for Colour $\suthree$} 
\label{action_SU3}

The standard Wilson action for the gluons is given by
\be
S^{\rm W}_G= \beta \sum_{\rm{sq}}\left[ 1 - \frac{1}{3}{\cal R}e
{\rm Tr} U_{\rm{sq}}(x)\right]
\label{Wilson_gaugeaction}
\ee
and a simple tree--level ${\cal O}(a^2)$--improved action (i.e., the
action with the first level of improvement) is defined as
\be
S_G=\frac{5\beta}{3}\sum_{\rm{sq}}{\cal R}e{\rm Tr}(1-U_{\rm{sq}}(x))-
\frac{\beta}{12u_{0}^2}\sum_{\rm{rect}}{\cal R}e{\rm Tr}
(1-U_{\rm{rect}}(x))\, .
\label{gaugeaction}
\ee
The $\obyo$ square (or plaquette) $U_{\rm{sq}}(x)$ and 
the $1\times 2$ rectangle $U_{\rm{rect}}(x)$ are 
defined by
\bea
U_{\rm{sq}}(x) & = & U_{\mu}(x)U_{\nu}(x+\hat{\mu})U^{\dagger}_{\mu}
(x+\hat{\nu})U^\dagger_{\nu}(x) \\
U_{\rm{rect}}(x) & = & U_{\mu}(x)U_{\nu}(x+\hat{\mu})U_{\nu}
(x+\hat{\nu}+\hat{\mu})U^{\dagger}_{\mu}(x+2\hat{\nu})
U^{\dagger}_{\nu}(x+\hat{\nu})U^\dagger_{\nu}(x) \nonumber \\
& + & U_{\mu}(x)U_{\mu}(x+\hat{\mu})U_{\nu}(x+2\hat{\mu})
U^{\dagger}_{\mu}(x+\hat{\mu}+\hat{\nu})U^{\dagger}_{\mu}
(x+\hat{\nu})U^\dagger_{\nu}(x).
\eea
Here the variables $\mu$ and $\nu$ are the direction in which 
the links are pointing inside
the lattice space. There are four directions for a 
four-dimensional hypercubic 
lattice. The link
product $U_{\rm{rect}}(x)$ denotes the rectangular $1\times2$
plaquettes and
$u_0$ is the tadpole improvement factor, commonly known as the 
mean--field improvement factor which largely corrects for 
quantum renormalization of the links. 
In our numerical studies we have typically employed
the plaquette definition of the mean--field improvement factor
\be
u_0=\left(\frac{1}{3}{\cal R}e{\rm Tr}\langle U_{\rm{sq}}\rangle
 \right)^{\frac{1}{4}}.
\label{uzero}
\ee
For the improved action in Eq.~(\ref{gaugeaction}) the residual
perturbative corrections after mean--field improvement are estimated
to be of the order of two to three percent~\cite{Alf95}.  Of course,
both Eqs.~(\ref{Wilson_gaugeaction}) and (\ref{gaugeaction}) reproduce
the continuum gluon action as $a\rightarrow{0}$, where $\beta\equiv
6/g^2$ and $g$ is the QCD coupling constant at the scale $a$.  It is
useful to note that our $\beta=6/g^2$ differs from the convention of
Refs.~\cite{Alf95,Lee,Woloshyn}.  A multiplication of our $\beta$ in
Eq.~(\ref{gaugeaction}) by a factor of $5/3$ reproduces their
definition.

Let us comment on the lattice configurations that we have generated
with the general algorithm described here and which we have used
extensively in Refs.~\cite{bonnet,bowman,zanotti,bowman2,bilson-thompson}.
The gauge configurations are generated using the 
Cabbibo-Marinari~\cite{Cab82} pseudo--heat--bath
algorithm with three diagonal $\sutwo$ subgroups.  All calculations
are performed using a highly parallel code written in CM-Fortran and
run on a Thinking Machines Corporations (TMC) CM-5 with 
appropriate link partitioning.
For the standard Wilson action we partition the link variable in a 
checkerboard fashion.  While all calculations to date have been for
$SU(3)$, there is no restriction in the algorithm on the number of colours
for the gauge group~\cite{Cab82} and we could just has easily have
treated the case of $SU(N)$.

The mean--field improvement factor was updated on a regular basis
during the simulation.  Once the lattice is thermalized from a cold
start, (after at least five thousand sweeps), the $u_0$ factor is held
fixed during the generation of the ensemble of gauge field
configurations.  The ensemble is built up by sampling the fields with
a separation of at least 500 Monte Carlo sweeps over the entire
lattice to ensure that they are sufficiently decorrelated.  For the
case of the standard Wilson action, configurations have been generated
on a $16^3 \times 32$ lattice at $\bt=5.70$ and a $24^3 \times 36$
lattice at $\bt=6.00$.  For the improved action of
Eq.~(\ref{gaugeaction}) we have generated $8^3 \times 16$, $12^3
\times 24$, $16^3 \times 32$, and $24^3 \times 36$ lattices with
$\beta$ values of 3.57, 4.10, 4.38, and 5.00 respectively.

\subsection{Masking and Parallel Computing}
\label{Masking}

When performing a Monte Carlo sweep of the entire lattice each lattice
link must be updated individually using the particular gluon action of
interest (e.g., $S_G$).  The action is used in the Monte Carlo
accept/reject step for that link in order that detailed balance is
ensured at each link update and hence that it is ensured throughout
the entire lattice sweep.  It is the combination of randomness in the
link updates, the maintenance of detailed balance, and decorrelation
(ensured by large sweep numbers between the taking of samples) that
ensures the desired ensemble of gauge field configurations are
produced with the Boltzmann distribution $\exp(-S_G)$.

In the most naive procedure we move through each link on the lattice
consecutively updating them one at a time until we have completed a
``sweep'' through the entire lattice.  We then repeat these lattice
sweeps as often as required.  This simple procedure is highly
inefficient on a parallel computing architecture, where we can be
updating many links at the same time.  However, there is a fundamental
limitation to this parallelism, i.e., we will violate detailed balance
and corrupt our data if we try to simultaneously update a link while
information about that link is being used in the update of another
link.  It is crucial that we identify which links can be updated
simultaneously and this is determined by the degree of non locality in
the action.  For example, for an action which contains only nearest
neighbor interactions of the links, such as the Wilson action, we can
use an efficient ``checkerboard'' algorithm, which will be described
below.  In general, the more non local is the lattice gluon action the
fewer are the links that can be simultaneously updated.  We see that
the improvement program is therefore more expensive to implement, but
the benefit of improved actions far outweighs this drawback.

In order to facilitate our discussions we will refer to the concept of
``masking'', where the lattice links not eliminated by the mask are
the ones that can be simultaneously updated in a parallel computing
environment.  The number of independent masks needed for a particular
action determines an upper limit to the parallelism that can be used
in a single lattice sweep.  As we will see, the best that can be done
is to have two masks per link direction and this is for the case of
nearest neighbor interactions only.

We will simplify the presentation in the usual way by rescaling all
dimensionful quantities by the lattice spacing $a$.  This is
equivalent to setting $a=1$.

\section{Masking for the Standard Wilson Action.}
\label{masking_Wilson}

In the standard Wilson action, where only neighbouring links are
connected by the action, we need only two masks for each of the four
link directions.  There are two different ways of implementing this
masking as we will now discuss.

\subsection{Checker Board Masking.}
\label{checkerboard}

The standard Wilson action only involves $\obyo$ Wilson loops (depicted 
in Fig.~\ref{plaquette}) and is the most fundamental lattice gluonic action. 
Whenever a given link is being updated, we must not be attempting
to update any of the links within any of the $1\times 1$ plaquettes
which contains the given link.  Consider the
link from the lattice site $x$ to $x+\mu$, where $\mu$ is one of the
four Cartesian unit vectors $\hat x, \hat y, \hat z$, or $\hat t$.
We see then that the plaquette in Fig.~\ref{plaquette} forms
a ``staple'' consisting of three links in the $\mu$-$\nu$
plane which is attached to the link of interest $U_{\mu}(x)$.
[Note that we are sometimes using $x$ as a shorthand notation
for the space-time lattice point $x^\mu \equiv (x,y,z,t)$
as well as for the $x$-coordinate on the $\hat x$ axis.  The
meaning should be clear from the context.]
We could equally well consider the plaquette and staple below
the link $U_{\mu}(x)$ in the figure, which also lies in the
$\mu$-$\nu$ plane.  In addition, for a given Cartesian direction
$\mu$, there are three possible choices for $\nu$, i.e., there
are three orthogonal planes which contain the link and two staples
per plane.  

\begin{figure}[hbt]
\centering{\
        \epsfig{angle=0,figure=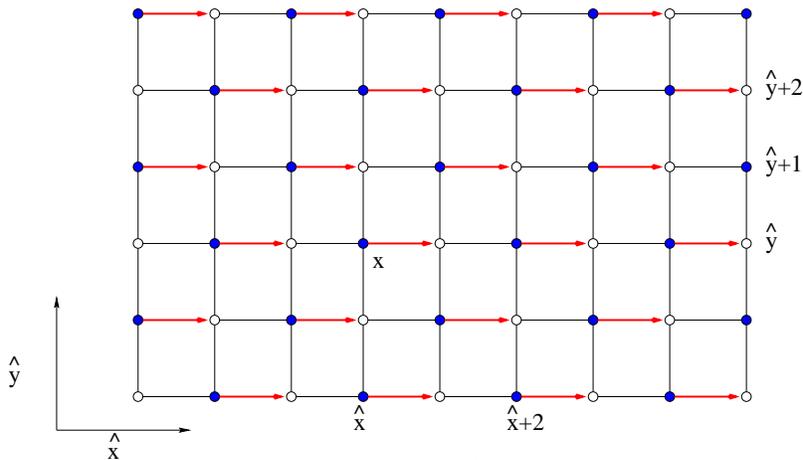,height=6cm} }
\parbox{130mm}{\caption{Checkerboard masking as seen in an
$\hat x$--$\hat y$ plane of the 
lattice when using the standard Wilson action. The highlighted
links with arrows can be updated simultaneously.}
\label{latticechek2}}
\end{figure}

Let us consider, for example, all of the links in the $\hat x$-$\hat y$
plane which are oriented in the $\hat x$ direction.  We can see from
Fig.~\ref{latticechek2} that we can choose a ``checkerboard'' of such
links that can be updated at the same time without interfering with
each other.  These links are indicated in the figure as
highlighted links with
arrows.   It is easy to see that none of the links to be updated lie in
any of the staples for the other links to be updated and that
exactly half of the $\hat x$-oriented links in this plane can be
simultaneously updated at one time.  

\begin{figure}[hbt]
\centering{\
        \epsfig{angle=0,figure=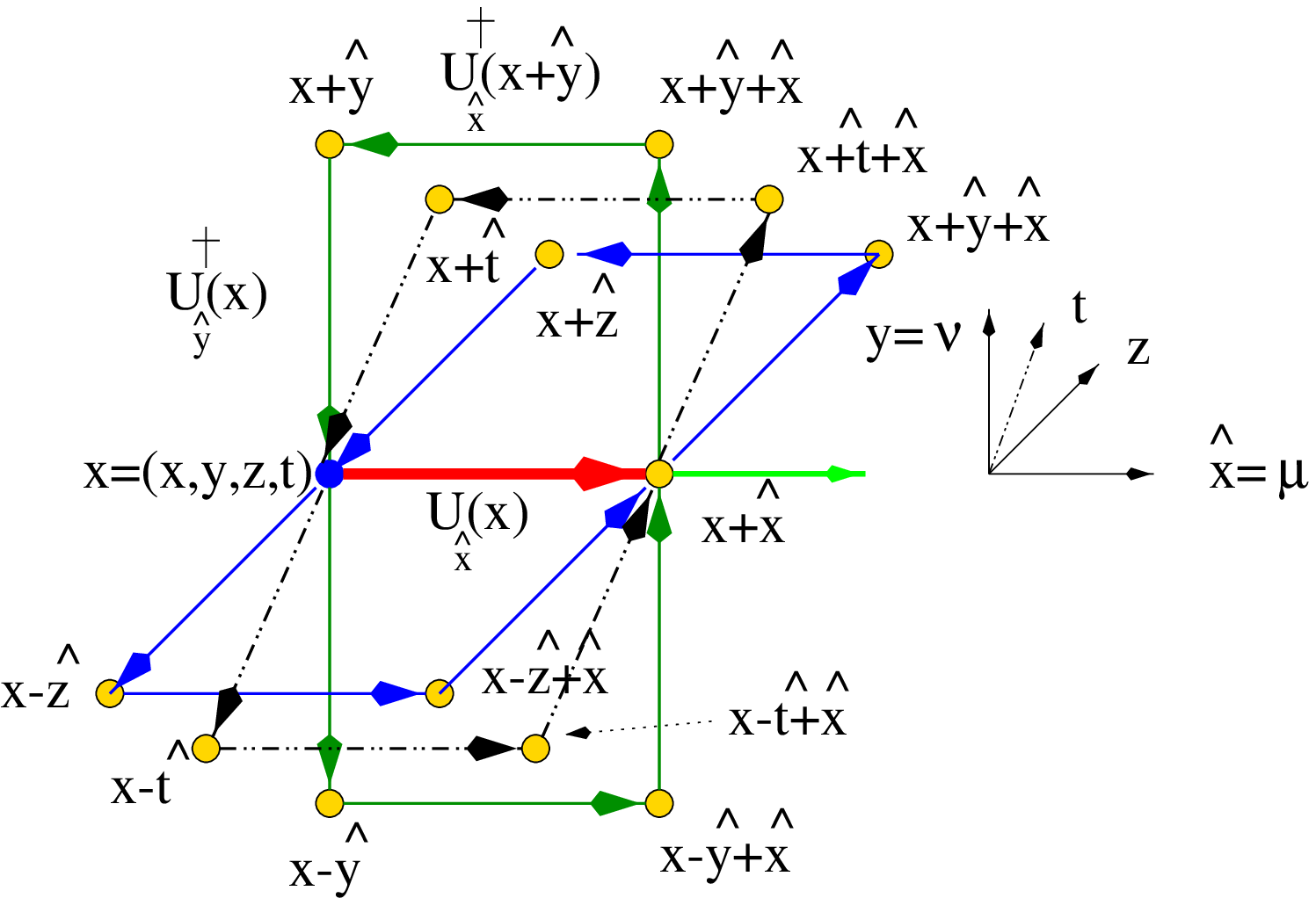,height=6cm} }
\parbox{130mm}{\caption{Rotating the $\obyo$ plaquette sitting in 
the $\xht\yht$ plane about the $\xht$--axis into the $\xht\zht$ and 
$\xht\tht$ planes.}
\label{staples}}
\end{figure}

We have identified one of the lattice sites in
Fig.~\ref{latticechek2}  as the site $x$.  If the
the link variable $U_{\xht}(x)$ is to updated
then from Fig.~\ref{latticechek2}, it is observed that the
link variables in the $\xht$ direction that can be
simultaneously updated are $U_{\xht}(x+2\xht)$,
$U_{\xht}(x+4\xht)$ and so 
on. So every second link along the $\hat x$ direction
can be updated at the same time. 
Now let us consider stepping in the $\hat y$ direction.
We again see that every second link in that direction
can be simultaneously updated.  By symmetry the same must also be
true for the $\hat z$ and $\hat t$ directions as depicted in
Fig.~\ref{staples}, where we have used a broken dash-dot line
to try to indicate the fourth dimension (i.e., for the links
that lie in the $\hat x$--$\hat t$ plane).  
We see that for the link pointing in the
$\hat x$ direction, the plaquettes (and staples) in the
$\hat x$--$\hat y$, $\hat x$--$\hat z$, $\hat x$--$\hat t$
planes are all related by simple rotations about the link. 
Thus we see that we have now built up a four-dimensional
mask for determining which links pointing in the $\hat x$ direction
can be simultaneously updated.

Let us introduce some convenient shorthand notation.  If for
a given link pointing in the direction $\mu$, we must take
$n$ steps in the direction $\nu$ before reaching the
next updatable link pointing in the direction $\mu$, we will
use the notation $\mu: \nu\sim n\nu$. 
For our checkerboard masking we see that for a link pointing
in the direction $\hat x$ we have to take two steps in 
each of the Cartesian directions before reaching the next updatable
link.  Hence we write
\bea
\hat x:\hspace*{0.5cm} 
\xht\sim{2\xht}\hspace*{0.25cm}{\rm ,}\hspace*{0.5cm}\yht\sim{2\yht}
\hspace*{0.25cm}{\rm ,}\hspace*{0.5cm}\zht\sim{2\zht}&\hspace*{0.25cm}
{\rm and}\hspace*{0.5cm}&\tht\sim{2\tht}.
\label{maskwilcheckrel}
\eea
We immediately see that this is also true for links oriented
in the $\hat y$, $\hat z$, and $\hat t$ directions so that 
\bea
\hat y:\hspace*{0.5cm} 
\xht\sim{2\xht}\hspace*{0.25cm}{\rm ,}\hspace*{0.5cm}\yht\sim{2\yht}
\hspace*{0.25cm}{\rm ,}\hspace*{0.5cm}\zht\sim{2\zht}&\hspace*{0.25cm}
{\rm and}\hspace*{0.5cm}&\tht\sim{2\tht}, \\
\hat z:\hspace*{0.5cm} 
\xht\sim{2\xht}\hspace*{0.25cm}{\rm ,}\hspace*{0.5cm}\yht\sim{2\yht}
\hspace*{0.25cm}{\rm ,}\hspace*{0.5cm}\zht\sim{2\zht}&\hspace*{0.25cm}
{\rm and}\hspace*{0.5cm}&\tht\sim{2\tht}, \\
\hat t:\hspace*{0.5cm} 
\xht\sim{2\xht}\hspace*{0.25cm}{\rm ,}\hspace*{0.5cm}\yht\sim{2\yht}
\hspace*{0.25cm}{\rm ,}\hspace*{0.5cm}\zht\sim{2\zht}&\hspace*{0.25cm}
{\rm and}\hspace*{0.5cm}&\tht\sim{2\tht}.
\eea

Finally, note that when we wish to update all of the links pointing
in any one of the four Cartesian directions, say $\mu$, we need only
two four-dimensional masks.  This is because
exactly half of the $\mu$-oriented links across the entire lattice
are considered in each four-dimensional mask.
To appreciate this we simply note that for any one of the Cartesian
directions one mask can be turned into the checkerboard complement
mask for that direction by shifting the mask by one step in any
Cartesian direction, (see Fig.~\ref{latticechek2}).
So to update all of the links on the
lattice we need a total of 8 four-dimensional masks,
i.e., 2 masks for each of
the four Cartesian directions.  In other words, no matter how
many nodes we have available on our parallel computing
architecture a full lattice updating sweep will require
8 serial masked sweeps to complete with a nearest neighbour action
(such as the Wilson action) and checkerboard masking.
This is the conventional procedure for the standard Wilson action
in lattice QCD studies.  In closing this section on the standard
Wilson action let us observe in Sec.~\ref{linearmask}
that there is an alternative
and equally good ``linear'' masking for this case.

\subsection{Linear Masking.}
\label{linearmask}

\begin{figure}[hbt]
\centering{\
        \epsfig{angle=0,figure=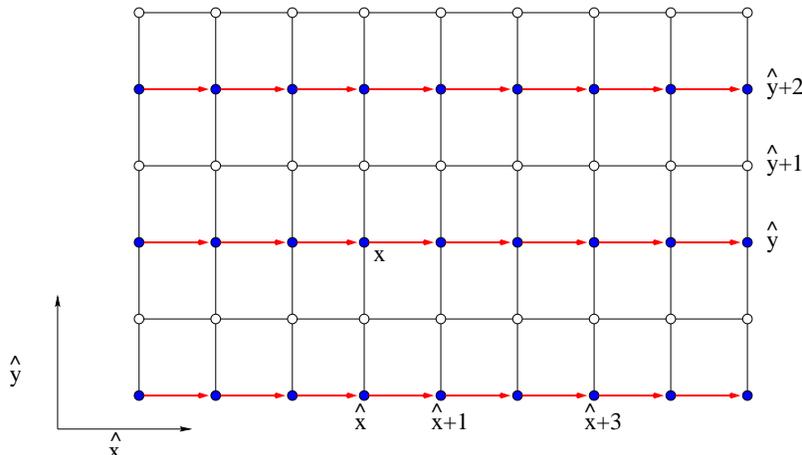,height=6cm} }
\parbox{130mm}{\caption{Linear masking of the lattice when using standard 
Wilson action. The highlighted arrows represents the link variable 
that can be updated simultaneously.}
\label{latticechek1}}
\end{figure}

As an alternative approach to the checker board masking described in 
Sec.~\ref{checkerboard}, one could
partition the links over the lattice in a linear fashion as shown in 
Fig.~\ref{latticechek1}.
If the link variable of interest is $U_{\xht}(x)$ then the next 
possible link variable in the
$\xht$ direction which can be updated is the $U_{\xht}(x+\xht)$ 
link and then the
$U_{\xht}(x+2\xht)$ and so on. We see that all the links on the 
$\xht$ line can be updated at the
same time, since none of these links are contained in the
$1\times 1$ plaquettes for the other links in the line.
Hence we have $\hat x: \hat x\sim 1\hat x$.
Now looking in the $\hat{y}$ 
direction, we realize that we cannot touch the
$U_{\xht}(x+\hat{y})$ link because it is part of the Wilson loop 
containing the link variable
$U_{\xht}(x)$ which is being updated simultaneously.
However, the links $U_{\xht}(x+2\hat{y})$,
$U_{\xht}(x+4\hat{y})$, etc.\ can be
updated. Consequently, we have
$\hat x: \hat{y}\sim{2\hat{y}}$ and similarly for steps
in the $\hat z$ and $\hat t$ directions.
For a link variable pointing in the $\xht$ direction
we then have that
\bea
\hat x: \hspace*{0.5cm}
\xht\sim{1\xht}\hspace*{0.25cm}{\rm ,}\hspace*{0.5cm}\yht
\sim{2\yht}\hspace*{0.25cm}{\rm ,}\hspace*{0.5cm}\zht\sim{2\zht}&
\hspace*{0.25cm}{\rm and}\hspace*{0.5cm}&\tht\sim{2\tht}.
\label{maskwilrel}
\eea
When the links to be updated are pointing in the other three
directions we have
\bea
\hat y:\hspace*{0.5cm}
\xht\sim{2\xht}\hspace*{0.25cm}{\rm ,}\hspace*{0.5cm}\yht\sim{1\yht}
\hspace*{0.25cm}{\rm ,}\hspace*{0.5cm}\zht\sim{2\zht}&\hspace*{0.25cm}
{\rm and}\hspace*{0.5cm}&\tht\sim{2\tht}, \\
\hat z:\hspace*{0.5cm}
\xht\sim{2\xht}\hspace*{0.25cm}{\rm ,}\hspace*{0.5cm}\yht\sim{2\yht}
\hspace*{0.25cm}{\rm ,}\hspace*{0.5cm}\zht\sim{1\zht}&\hspace*{0.25cm}
{\rm and}\hspace*{0.5cm}&\tht\sim{2\tht}, \\
\hat t:\hspace*{0.5cm}
\xht\sim{2\xht}\hspace*{0.25cm}{\rm ,}\hspace*{0.5cm}\yht\sim{2\yht}
\hspace*{0.25cm}{\rm ,}\hspace*{0.5cm}\zht\sim{2\zht}&\hspace*{0.25cm}
{\rm and}\hspace*{0.5cm}&\tht\sim{1\tht},
\eea
for the $\yht,\zht$ and $\tht$ directions respectively. 

Again, we see that there are two complementary linear masks for
links pointing in any given Cartesian direction $\mu$.  One mask can
be obtained from the other by a shift of one step in any
of the three Cartesian directions orthogonal to $\mu$ as
can be appreciated from Fig.~\ref{latticechek1}.
Thus this linear masking is equally as efficient as the
checkerboard masking of the previous section, since there
are 2 masks for each of the 4 Cartesian directions giving a total
of 8 masks.

\section{Masking  an Improved Action.}
\label{masking_improved}

\begin{figure}[hbt]
\centering{\
        \epsfig{angle=0,figure=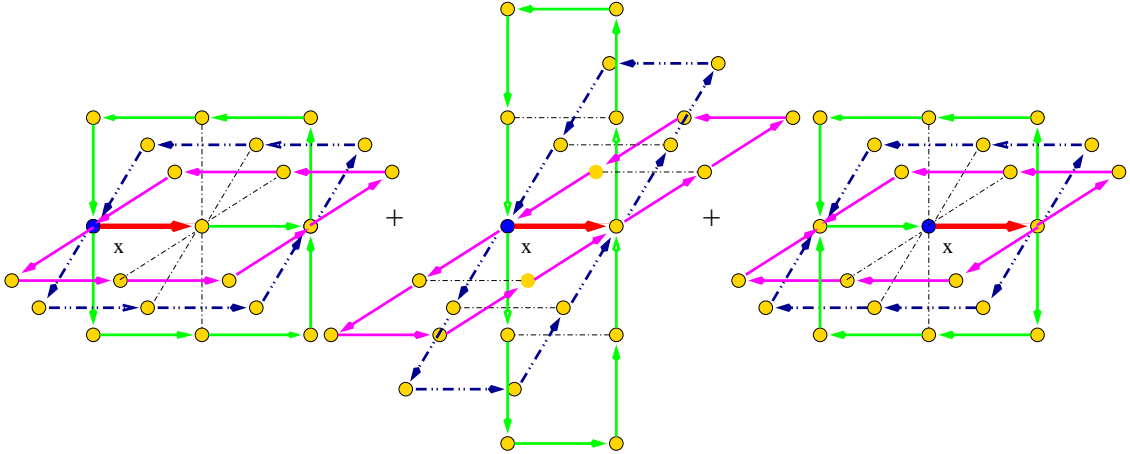,height=6cm}}
\parbox{130mm}{\caption{The set of all possible $1\times 2$ plaquettes
containing the link $U_{\hat x}(x)$.  The dashed-dotted line is to
be understood as being in the $\hat x$--$\hat t$ plane.}
\label{fullrectangles}}
\end{figure}

In this section, we describe the necessary masking procedure for 
a first-level improved action involving $1\times 1$ and
$1\times 2$ Wilson loops.  In particular, in this section
we are describing the masking suitable for the improved gauge
action of Eq.~(\ref{gaugeaction}), which has been used
extensively by us~\cite{bonnet,bowman,zanotti,bowman2,bilson-thompson}. 
Let us again begin by considering the link variable beginning at
some lattice site $x$ and pointing in the $\hat x$ direction, i.e.,
$U_{\hat x}(x)$.
We now need to consider both Fig.~\ref{staples} for the
elementary $1\times 1$ square plaquette and Fig.~\ref{fullrectangles}
for the $1\times 2$ rectangular plaquette.  In Fig.~\ref{fullrectangles}
we show all of the $1\times 2$ rectangular plaquettes which contain 
the link $U_{\hat x}(x)$, which is shown as the highlighted horizontal 
link in the three parts of this figure. Visualizing a
four dimensional object on a flat piece of paper can be, to a certain
extent, an artistic challenge and so we have again used a dash--dot
line to indicate links lying in the $\hat x$--$\hat t$ plane.
There are three distinguishable ways to include this link in a
$1\times 2$ plaquette (the three parts of the figure)
and for each of these there are two (mirror-image) rectangles per
Cartesian plane and four Cartesian planes.
All links in Figs.~\ref{staples} and \ref{fullrectangles}
with arrows (other than the link $U_{\hat x}(x)$ itself)
must be omitted from the mask when updating this link with
our improved action.  We see that there are many excluded links.

\begin{figure}[hbt]
\centering{\
        \epsfig{angle=0,figure=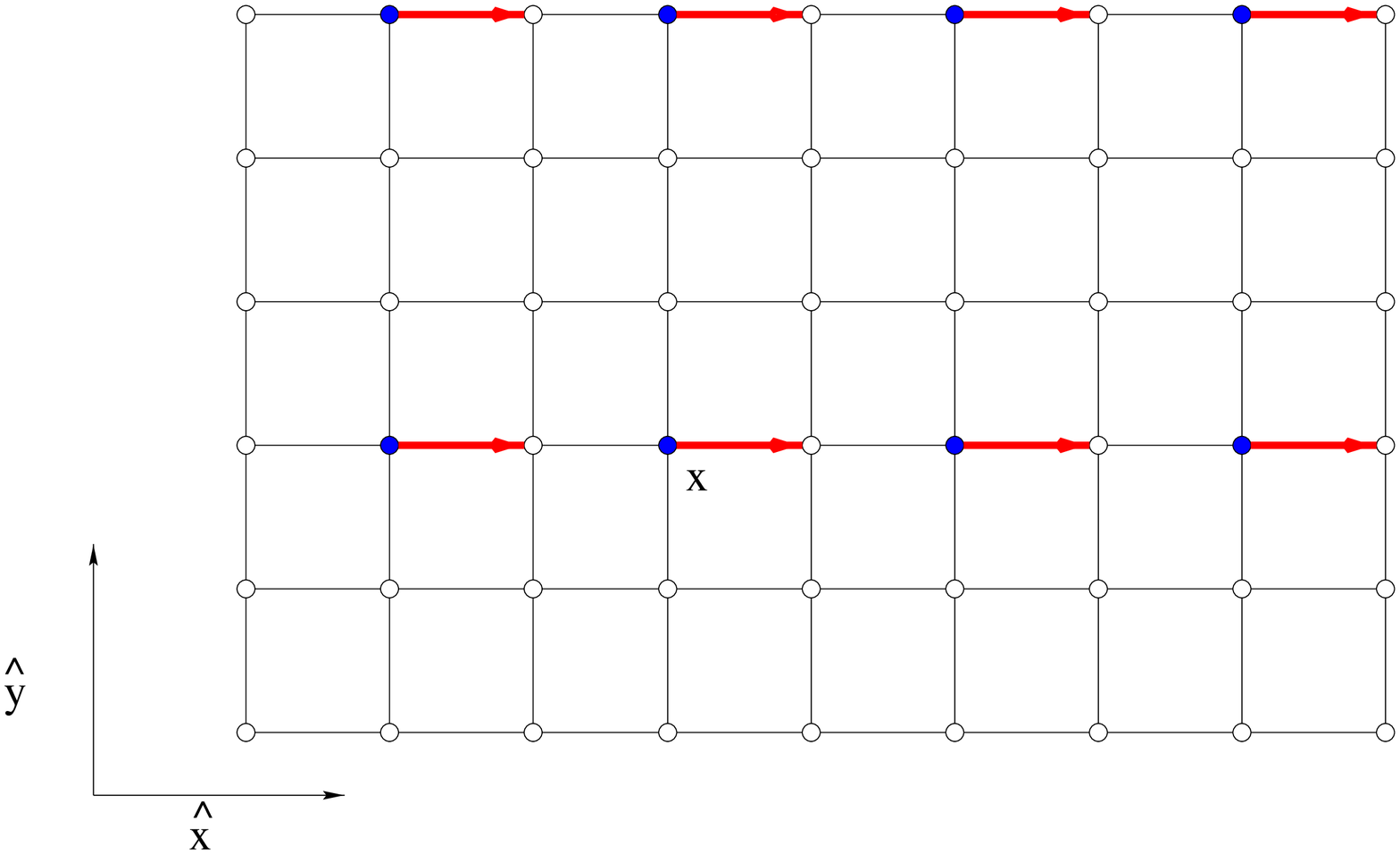,height=6cm} }
\parbox{130mm}{\caption{The highlighted links with arrows are the ones
that can be simultaneously updated for an action containing both
$1\times 1$ and $1\times 2$ plaquettes.
}
\label{lattice}}
\end{figure}

In Fig.~\ref{lattice} we show which links can
be simultaneously updated with the link $U_{\hat x}(x)$.  We can
immediately write down by inspection from this figure that
\bea
\hat x:\hspace*{0.5cm}
\xht\sim{2\xht}\hspace*{0.25cm}{\rm ,}\hspace*{0.5cm}\yht\sim{3\yht}
\hspace*{0.25cm}{\rm ,}\hspace*{0.5cm}\zht\sim{3\zht}&\hspace*{0.25cm}
{\rm and}\hspace*{0.5cm}&\tht\sim{3\tht}\, .
\label{maskimprel}
\eea
This follows since the $\hat z$ and $\hat t$ cases are identical
to the $\hat y$ case for this $\hat x$--oriented link. 
The generalization to the other orientations of the links to be
updated is straightforward by symmetry
\bea
\hat y:\hspace*{0.5cm}
\xht\sim{3\xht}\hspace*{0.25cm}{\rm ,}\hspace*{0.5cm}\yht\sim{2\yht}
\hspace*{0.25cm}{\rm ,}\hspace*{0.5cm}\zht\sim{3\zht}&\hspace*{0.25cm}
{\rm and}\hspace*{0.5cm}&\tht\sim{3\tht}, \\
\hat z:\hspace*{0.5cm}
\xht\sim{3\xht}\hspace*{0.25cm}{\rm ,}\hspace*{0.5cm}\yht\sim{3\yht}
\hspace*{0.25cm}{\rm ,}\hspace*{0.5cm}\zht\sim{2\zht}&\hspace*{0.25cm}
{\rm and}\hspace*{0.5cm}&\tht\sim{3\tht}, \\
\hat t:\hspace*{0.5cm}
\xht\sim{3\xht}\hspace*{0.25cm}{\rm ,}\hspace*{0.5cm}\yht\sim{3\yht}
\hspace*{0.25cm}{\rm ,}\hspace*{0.5cm}\zht\sim{3\zht}&\hspace*{0.25cm}
{\rm and}\hspace*{0.5cm}&\tht\sim{2\tht}\, .
\eea

Let us return to the particular case of the masking for
$\hat x$-oriented links.  From Eq.~(\ref{maskimprel}) we
see that there is symmetry between the $\hat y$, $\hat z$,
and $\hat t$ directions and so we will begin by constructing
suitable masks for any given equal-$x$ hyper-plane, i.e., for the
three dimensional space spanned by the unit vectors 
$\hat y$, $\hat z$, and $\hat t$.

\begin{figure}[hbt]
\centering{\
        \epsfig{angle=0,figure=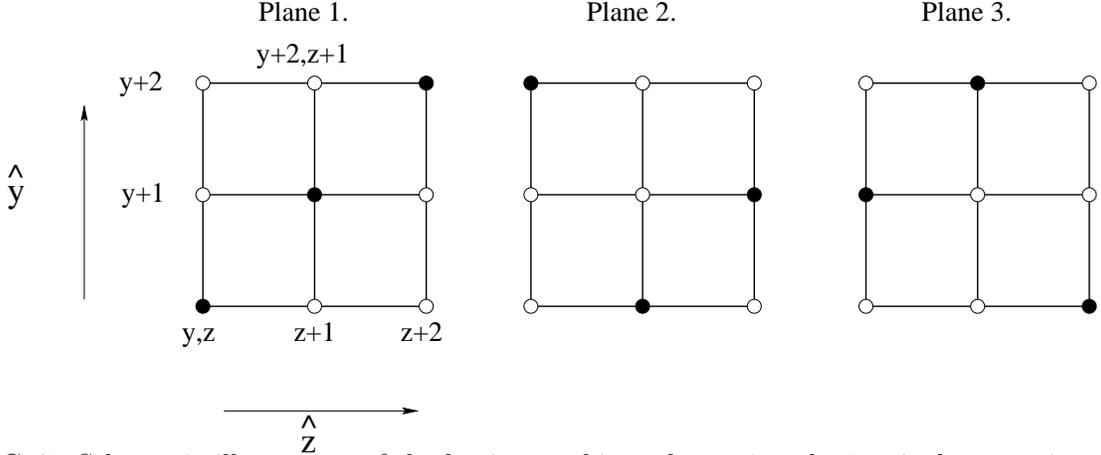,height=6cm} }
\parbox{130mm}{\caption{Schematic illustration of the lattice masking when
using the $1\times 2$ plaquette improved action.}
\label{impmask}}
\end{figure}

Before attempting this, let us first consider Fig.~\ref{lattice}
and extend this to three dimensions
by imagining that the $\hat z$-axis is pointing directly out from the page.
We shall temporarily neglect the $\hat t$ direction, which is
equivalent to simply taking a slice of the four-dimensional lattice
with the same value of $t$, (i.e., an equal-$t$ hyper-plane).
Now let us view this three-dimensional lattice by looking along 
the $\hat x$-axis at one particular equal-$x$ plane.  We will then be
presented with end-views of updatable links in the $\hat y$--$\hat z$
plane.  For every fixed value of $z$ there are three different masks
needed for $y$ and vice versa.  Also, there is no restriction
on simultaneously updating {\em diagonally} shifted links, since we
are only considering planar actions at this point.  It is not difficult
to see that we can cover all of the nine lattice links that
need to be updated with three orthogonal masks as shown in
Fig.~\ref{impmask}.  In this figure $x$-oriented links which can be
updated at the same time are indicated by a solid dot.  Note that each of
these masks is related by a diagonal shift of the nine-point
lattice ``window''.  

\begin{figure}[hbt]
\centering{\
        \epsfig{angle=0,figure=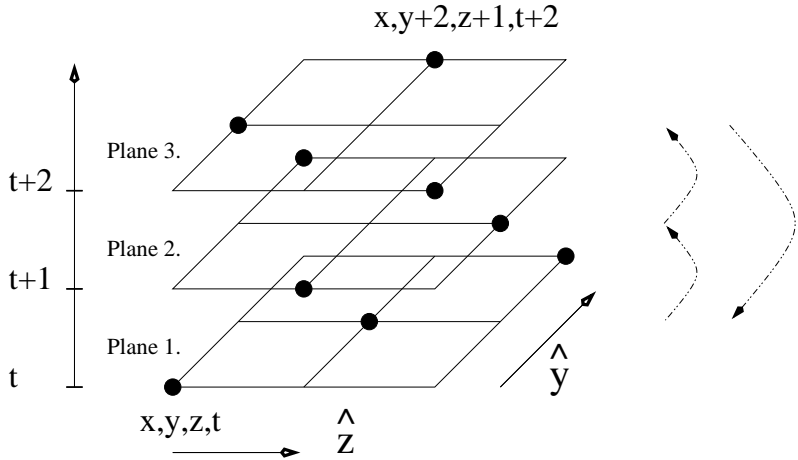,height=6cm} }
\parbox{130mm}{\caption{Illustration of the cyclic plane rotation in the
improved masking.}
\label{planes}}
\end{figure}

We can now also extend this thinking to include the $t$ direction,
by stacking the three two-dimensional $y$--$z$ masks on top of each other
as shown in Fig.~\ref{planes}.  We must stack the planes so that
when viewed along any of the three axes the solid dots in any one
Cartesian planes always have the appearance of one of the planes
in Fig.~\ref{impmask}.  We see that this can be achieved in three
ways by the stacking in Fig.~\ref{planes} and its two cyclic permutations.
These three three-dimensional masks when summed give the identity
(i.e., the sum includes all points) and are orthogonal to each other
(i.e., the sum includes all points only once).

We can now give a simple geometrical picture of what we are doing,
which will simplify the generalization that we give in the next section.
For $\hat x$-oriented links, the directions $\hat y$, $\hat z$,
and $\hat t$ directions are all symmetrical and each direction requires
a step of 3 to reach the next updatable link.  Hence, we need
to construct a complete set of orthogonal masks in three dimensions for a
$3\times 3\times 3$ cube, where no two points in the cube lie on the
same Cartesian axis (i.e., only diagonally related points).
This is simple to do.  Let us consider 
the bottom plane (i.e., plane 1) of Fig.~\ref{planes} and connect the
three solid dots by a diagonal line.  We see that plane 2 is obtained
from plane 1 by a diagonal shift of this line by one diagonal half-step,
and similarly for plane 3.  In visualizing this it may help to imagine
surrounding the cube by many identical copies of itself and moving the
diagonal line through diagonal half-steps across all of these cubes
simultaneously.  All three three-dimensional masks are obtained in
the same way but start with plane 1, plane 2, and plane 3 respectively.  
 
So for the $\hat x$-oriented links we need 3 masks for each equal-$x$
hyper-plane (i.e., a three-volume here) and we have two independent 
equal-$x$ hyper-planes, giving a total of
6 masks for each Cartesian direction for the link orientation.
Since there are 4 orientations, then there is a total of 24 masks
needed for an action containing both $1\times 1$ and $1\times 2$
plaquettes.  Thus a single lattice sweep must take at least 24 sequential
serial calculations even on the most parallel computing architecture.  
 
The masking procedure outlined here for this action can only be
implemented when the number of lattice points in each dimension is a
multiple of three.  Inspection of Fig.~\ref{planes} reveals the
periodicity of three is required to maintain separation of links at
the boundary.  Since simulations are usually carried out on lattices
with even numbered sides, this restricts the length of the lattice
sides to multiples of six.  Fortunately, multiples of four are easily
obtained as described in the next section.  Moreover,
Sec.~\ref{nonplanar} reports a high-performance mask for this action
with a periodicity of four.

It is interesting to note that when implementing this masking
procedure on the CM-5 we achieved optimum performance by calculating
the updates for all links on the lattice and by then only implementing
those updates that were appropriate for the particular mask being used
at the time.  In other words for the lattices that we have studied so
far on the CM-5 it was more efficient to calculate link updates that
were never used, than it was to split the masked links over the
various processor nodes and update only these masked links.  This was
due to the fact that there was a large overhead of communication time
in assigning the masked links across the processors.  The point of
this observation is that the optimal use of the masks will in general
depend on the details of the parallel computing architecture being
used.

\section{Masking the Lattice When Using a Generalized Improved Action.}
\label{maskingsupimpsu3}

\begin{figure}[hbt]
\centering{\
        \epsfig{angle=0,figure=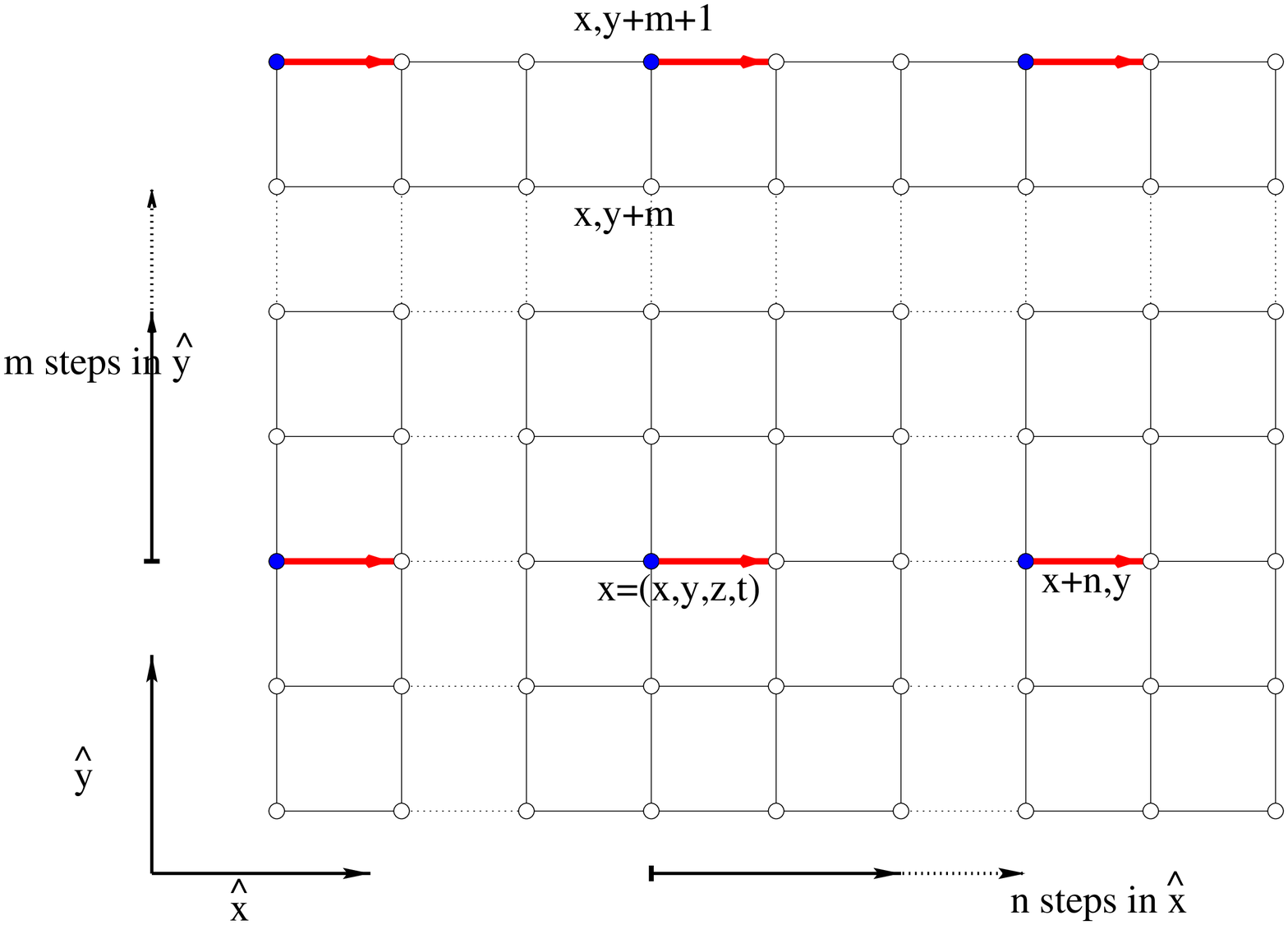,height=7cm} }
\parbox{130mm}{\caption{
The highlighted links with arrows are the ones
that can be simultaneously updated for an action containing
up to $n\times m$ plaquettes, where here $n$ refers to the $\hat x$
direction and $m$ applies to the other three Cartesian directions.
}
\label{nbymlattice}}
\end{figure}

We can now generalize the algorithm presented in
Sec.~\ref{masking_improved} for arbitrarily improved planar actions.
Let us begin as before by considering the update of links oriented in
the $\hat x$-direction.  Let us assume that we have an action with
$n\times m$ links where the $n$ refers to the $\hat x$ direction and
the $m$ refers to the $\hat y$, $\hat z$, $\hat t$ directions.  We
will eventually argue that only the $n_{\rm max}\times n_{\rm max}$
case, where $n_{\rm max}$ is the greater of $n$ and $m$, is necessary
in the general case.  As shown in Fig.~\ref{nbymlattice} the nearest
simultaneously updatable links are separated by $n$ steps in the $\hat
x$ direction and $(m+1)$ steps in the other three Cartesian
directions.

Hence we see that we can write in our notation for the four Cartesian
orientations of the links that
\bea
&&\hat x:\hspace*{0.5cm}
\xht\sim{n\xht}\hspace*{0.25cm}{\rm ,}\hspace*{0.5cm}\yht\sim{(m+1)\yht}
\hspace*{0.25cm}{\rm ,}\hspace*{0.5cm}\zht\sim{(m+1)\zht}\hspace*{0.25cm}
{\rm and}\hspace*{0.5cm}\tht\sim{(m+1)\tht}, \\
&&\hat y:\hspace*{0.5cm}
\xht\sim{(m+1)\xht}\hspace*{0.25cm}{\rm ,}\hspace*{0.5cm}\yht\sim{n\yht}
\hspace*{0.25cm}{\rm ,}\hspace*{0.5cm}\zht\sim{(m+1)\zht}\hspace*{0.25cm}
{\rm and}\hspace*{0.5cm}\tht\sim{(m+1)\tht}, \\
&&\hat z:\hspace*{0.5cm}
\xht\sim{(m+1)\xht}\hspace*{0.25cm}{\rm ,}\hspace*{0.5cm}\yht\sim{(m+1)\yht}
\hspace*{0.25cm}{\rm ,}\hspace*{0.5cm}\zht\sim{n\zht}\hspace*{0.25cm}
{\rm and}\hspace*{0.5cm}\tht\sim{(m+1)\tht}, \\
&&\hat t:\hspace*{0.5cm}
\xht\sim{(m+1)\xht}\hspace*{0.25cm}{\rm ,}\hspace*{0.5cm}\yht\sim{(m+1)\yht}
\hspace*{0.25cm}{\rm ,}\hspace*{0.5cm}\zht\sim{(m+1)\zht}\hspace*{0.25cm}
{\rm and}\hspace*{0.5cm}\tht\sim{n\tht}\, .
\eea

\begin{figure}[hbt]
\centering{\
        \epsfig{angle=0,figure=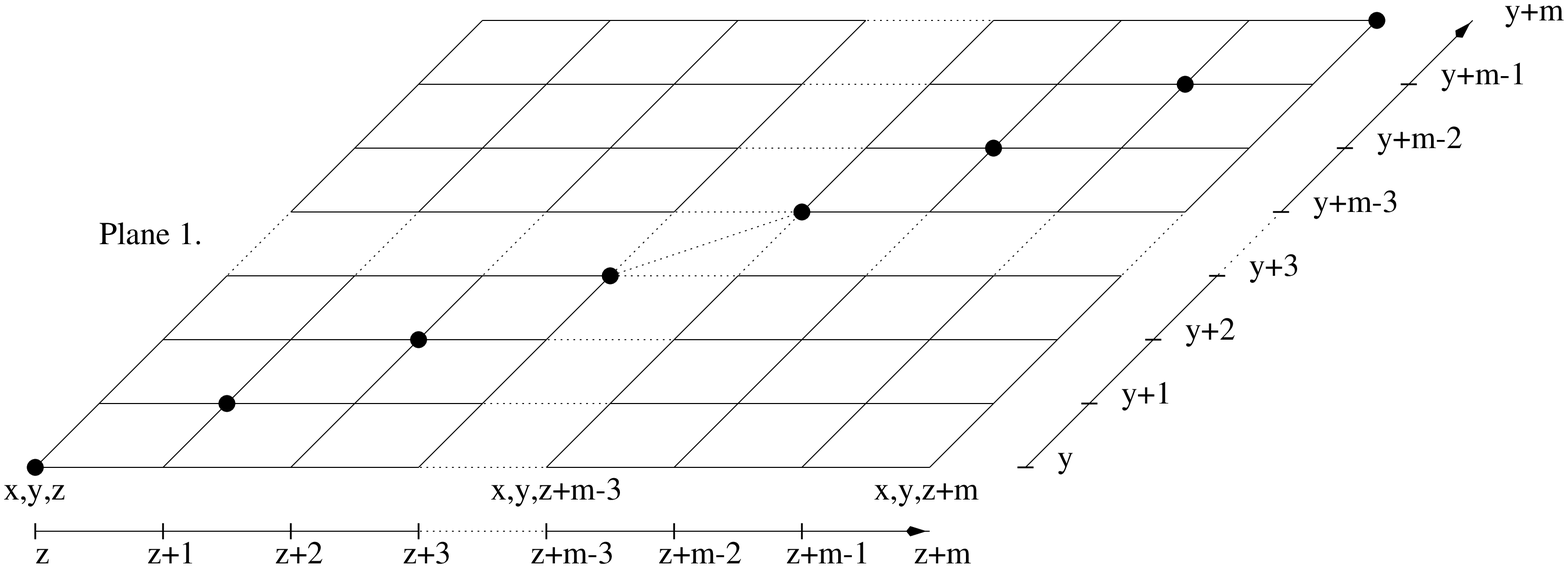,height=5cm}}
\parbox{130mm}{\caption{Plane 1 with the $(m+1)$ updatable sites 
on the main diagonal of the $\yht,\zht$ plane.}
\label{nbymplanes1that}}
\end{figure}

\begin{figure}[hbt]
\centering{\
        \epsfig{angle=0,figure=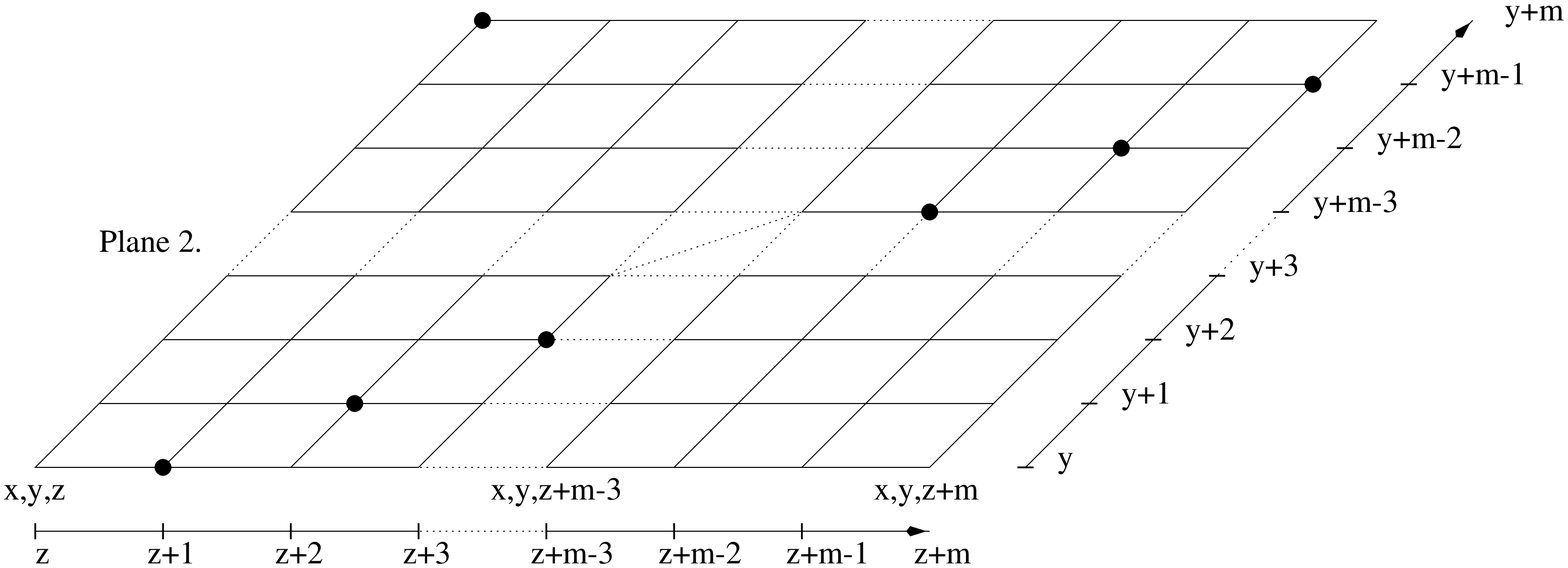,height=5cm}}
\parbox{130mm}{\caption{Plane 2.}
\label{nbymplanes2thatp1}}
\end{figure}

\begin{figure}[hbt]
\centering{\
        \epsfig{angle=0,figure=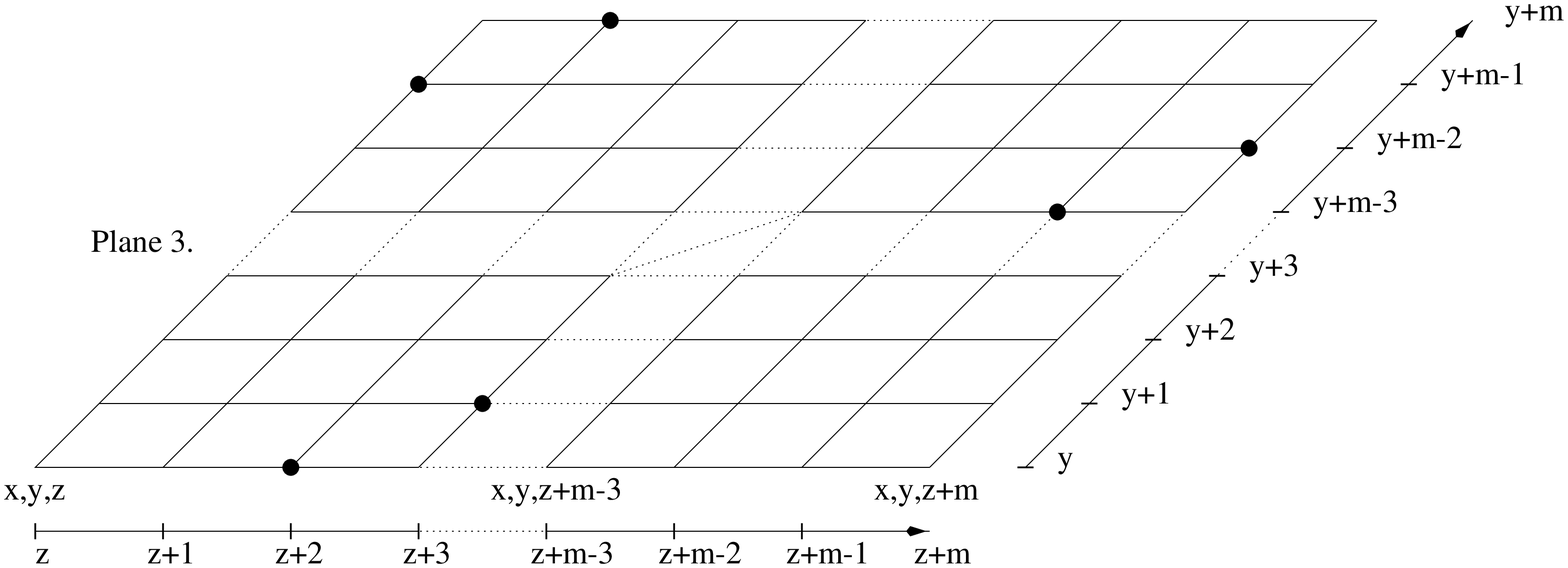,height=5cm}}
\parbox{130mm}{\caption{Plane 3.}
\label{nbymplanes3thatp2}}
\end{figure}

We can now follow the arguments of the previous section.  Let us consider
a fixed-$x$ hyper-plane (i.e., three-volume).  In place of a 
$3\time 3\times 3$ three-volume we will now need an 
$(m+1)\times (m+1)\times (m+1)$
three-volume.  Furthermore, we will need a complete set of orthogonal
and diagonal masks for this.  
Let us again look along the $\hat x$ direction at a fixed $t$ plane
for now, i.e., we are looking at a $\hat y$--$\hat z$ plane
as in Fig.~\ref{nbymplanes1that}.  Let us refer to the 
$(m+1)\times (m+1)$ two-dimensional
plane with the updatable links (solid dots) along the diagonal as
plane 1.  Then we can generate the other $m$ two-dimensional planes
by diagonal half-shifts as before as depicted in 
Figs.~\ref{nbymplanes2thatp1} and \ref{nbymplanes3thatp2}.
We can then sequentially stack these planes in the $\hat t$ direction
as before to form the first of the three-dimensional masks.
The other $m$ three-dimensional masks are then generated from this
first mask by the cyclic permutations of the $m+1$ planes as in
Sec.~\ref{masking_improved}.  Hence we have generated
the desired complete set of $(m+1)$ orthogonal three-dimensional
diagonal masks.


So for each fixed $x$-hyper-plane (i.e., three volume) we need $(m+1)$
masks.  We will need such a set of masks for the $n$ values of $x$.
The general result is that for updating the links oriented in the
$\hat x$ direction we need a total of $n\times(m+1)$ masks and we have
seen that the construction of these masks is straightforward.  The
construction of the masks for the other Cartesian orientations of the
links proceeds identically.  This total number of masks is $n_{\rm
mask}=4\times n \times (m+1)$.
The periodicity of the mask is governed by the last factor, $(m+1)$,
and the lengths of the lattice dimensions must be a multiple of this
number.  The reason for this is that if this were not the case then
the imposition of the necessary periodic boundary conditions would
cause link collisions, where a link being updated uses one or
more other links which are simultaneously being updated.

Any improved lattice action of physical interest must be both
$Z_4$-symmetric (i.e., symmetric under the arbitrary interchange of
the four Cartesian directions) and translationally invariant.
Thus for such actions every link will find itself occurring in every
possible position for every plaquette in the improved action.  We then
see, as we did in Sec.~\ref{masking_improved} and
Fig.~\ref{fullrectangles}, that the number of steps needed in each
direction is determined by the longest plaquette side appearing in the
action.  Let us denote the longest plaquette side appearing in the
action as $n_{\rm max}$.  Then we see that the number of steps needed
in the various Cartesian directions is given by
\bea
&&\hat x:\hspace*{0.5cm}
\xht\sim{n_{\rm max}\xht}\hspace*{0.25cm}{\rm ,}
\hspace*{0.5cm}\yht\sim{(n_{\rm max}+1)\yht}
\hspace*{0.25cm}{\rm ,}\hspace*{0.5cm}\zht\sim{(n_{\rm max}+1)\zht}
\hspace*{0.25cm}
{\rm and}\hspace*{0.5cm}\tht\sim{(n_{\rm max}+1)\tht}, \\
&&\hat y:\hspace*{0.5cm}
\xht\sim{(n_{\rm max}+1)\xht}\hspace*{0.25cm}{\rm ,}\hspace*{0.5cm}
\yht\sim{n_{\rm max}\yht}
\hspace*{0.25cm}{\rm ,}\hspace*{0.5cm}\zht\sim{(n_{\rm max}+1)\zht}
\hspace*{0.25cm}
{\rm and}\hspace*{0.5cm}\tht\sim{(n_{\rm max}+1)\tht}, \\
&&\hat z:\hspace*{0.5cm}
\xht\sim{(n_{\rm max}+1)\xht}\hspace*{0.25cm}{\rm ,}\hspace*{0.5cm}
\yht\sim{(n_{\rm max}+1)\yht}
\hspace*{0.25cm}{\rm ,}\hspace*{0.5cm}
\zht\sim{n_{\rm max}\zht}\hspace*{0.25cm}
{\rm and}\hspace*{0.5cm}\tht\sim{(n_{\rm max}+1)\tht}, \\
&&\hat t:\hspace*{0.5cm}
\xht\sim{(n_{\rm max}+1)\xht}\hspace*{0.25cm}{\rm ,}\hspace*{0.5cm}
\yht\sim{(n_{\rm max}+1)\yht}
\hspace*{0.25cm}{\rm ,}\hspace*{0.5cm}\zht\sim{(n_{\rm max}+1)\zht}
\hspace*{0.25cm}
{\rm and}\hspace*{0.5cm}\tht\sim{n_{\rm max}\tht}\, .
\eea
Hence the number of masks in general for an improved action will then
be given by
\be
n_{\rm mask} = 4\times n_{\rm max}\times (n_{\rm max}+1)
\label{Eq:n_mask}
\ee 
and the lattice will need the length in each dimension to be an integral
multiple of $(n_{\rm max}+1)$.

It is useful to note that the {\em linear} masking for the standard
Wilson action is the one that is extended initially in
Sec.~\ref{masking_improved} and is subsequently generalized in this
section.  For the standard Wilson action (i.e., $1\times 1$ plaquettes
only) we see that $n_{\rm max}=1$ and hence $n_{\rm mask}=4\times
1\times 2= 8$ as we found for the linear (and checkerboard) mask.  For
the improved action that we have studied (i.e., $1\times 1$ and
$1\times 2$ plaquettes) we have $n_{\rm max}=2$ and hence $n_{\rm
mask}=4\times 2\times 3= 24$ or 6 masks per link direction as found in
Sec.~\ref{masking_improved}.  However, this way of proceeding for the
plaquette plus rectangle improved action would require each lattice
dimension be a multiple of $(n_{\rm max}+1)=3$, but since we also
typically want our lattices to have even lengths then that means
each side of the lattice would need to be a multiple of 6 in length.
Since the result in Eq.~(\ref{Eq:n_mask}) is a lower bound, we can of
course always choose to enlarge the period of our masking by
choosing $n_{\rm max}+2$ for the last factor in Eq.~\ref{Eq:n_mask}
rather than $n_{\rm max}+1$.  This will still ensure that no link
collisions occur.  For example, for the plaquette plus rectangle
improved action we can use $(n_{\rm max}+2)=4$ instead
of $(n_{\rm max}+1)=3$ in Eq.~(\ref{Eq:n_mask}), so that
any lattice lengths which are multiples
of 4 become available at the cost of requiring 32 masks rather than 24.
Fortunately, for this case a more efficient mask can be realized and
will be presented in the next section.

\section{Non-planar Considerations}
\label{nonplanar}

We have presented a method for identifying links which may be
simultaneously updated during Monte-Carlo updates or cooling sweeps.
The generality of the algorithm allows one to parallelize link updates
for planar actions of any degree of non locality.  In this section we
extend this analysis to a few special cases of actions in which
out-of-plane considerations are necessary.  Both cases are centred
around the plaquette plus rectangle action of Eq.~(\ref{gaugeaction}) in
which $1\times 1$ and $1\times 2$ Wilson loops are considered in the
action.  Such actions dominate current improved gauge action analyses.

In Sec.~\ref{masking_improved} we illustrated how such an action can
be masked through the consideration of an elementary $3 \times 3
\times 3$ cube in which one-third of the links may be simultaneously
updated.  However, only every second link in the direction of the
links is updated simultaneously as illustrated in
Fig.~\ref{lattice}.  Hence six masks per link direction are
required.   

\begin{figure}[htb]
\centering{\
        \epsfig{angle=0,figure=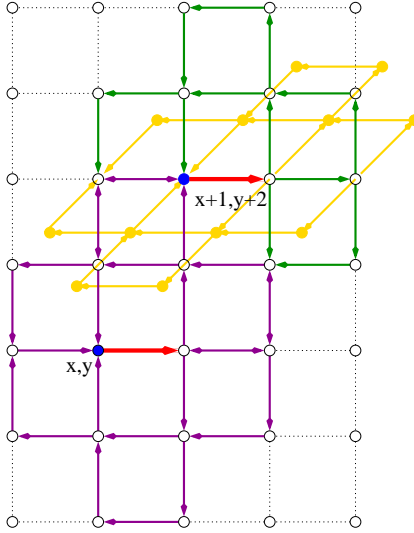,height=7cm} 
\vspace{12pt} 
}
\parbox{130mm}
{\caption{
Two elementary cells for an action involving $1\times 1$ and $1\times
2$ Wilson loops are nested together such that one need not restrict
the mask to every second link in the direction of the links being
updated.  The links with the positions labeled are the ones that can
be simultaneously updated.  The out of plane plaquette-plus-rectangle
illustrates additional links that cannot be simultaneously updated.
}
\label{1by2latticeNested}}
\end{figure}

\begin{figure}[htb]
\centering{\
        \epsfig{angle=0,figure=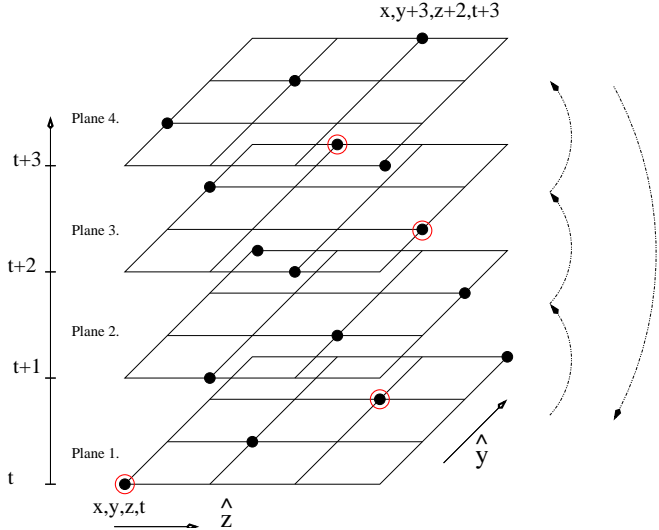,height=7cm} 
\vspace{12pt} 
}
\parbox{130mm}{\caption{
 The four planes to be cycled through in the elementary $4 \times 4
\times 4$ cube.  One-quarter of the links may be updated
simultaneously and are indicated by the solid dot.  The circled sites
are an example of the sites surviving when the out of plane ``chair''
or ``parallelogram'' link paths are included in the action.}
\label{1by2planes4select}}
\end{figure}

Here we consider an alternative masking specialized to the $1\times 1$
and $1\times 2$ Wilson loop actions.  Fig.~\ref{1by2latticeNested}
illustrates the manner in which these Wilson loops may be nested, such
that one need not restrict the mask to every second link in the
direction of the links being updated.  This technique will reduce the
number of masks by a factor of two, at the expense of considering an
elementary $4 \times 4 \times 4$ cube in which one-quarter of the
links may be simultaneously updated.  Fig.~\ref{1by2planes4select}
displays the four planes to be cycled through in which the links to be
updated simultaneously are indicated by the solid dot.  Hence only
four masks per link direction are required.  Moreover, the lattice
dimensions (usually even numbers) can now be multiples of four as
opposed to six.

The out-of-plane considerations required for the nested action are
also indicated in Fig.~\ref{1by2latticeNested}.  Hence it becomes
apparent that not only the three links at $(x,y+1)$, $(x,y+2)$, and
$(x,y+3)$, be avoided, but also the links two-steps in a direction
orthogonal to the link direction and one step in a third direction
(similar to moves of a Knight on a chess board) must be avoided.  

Inspection of the four planes to be cycled through in the elementary
$4 \times 4 \times 4$ cube displayed in Fig.~\ref{1by2planes4select}
indicates that such Knight moves are already avoided in this mask.
However, it also becomes clear that the ordering of the planes is
crucial.  For example interchanging the positions of planes 2 and 3
would cause ``link collisions'' within the nested mask.

Finally we consider non-planar actions in which one step out of the
plane of the $1\times 1$ and $1\times 2$ Wilson loops is required.
Such non-planar paths are introduced to eliminate small but finite
${\cal O}(g^2 a^2)$ errors where $g$ is the gauge coupling constant.
The six-link paths commonly referred to as the ``chair'' and
``parallelogram'' \cite{Alf95} introduce a link parallel to that being
updated which is one-step orthogonal to the link direction and one
step in a third direction.  

Inspection of Fig.~\ref{1by2planes4select} indicates that such 1 by 1
moves eliminates fully two of the four planes and half of the parallel
sites on each surviving plane.  An example of four of the sites which
may still be updated in parallel are indicated by the circled sites in
Fig.~\ref{1by2planes4select}.  As a result there are now 16 masks
required per link direction instead of 4.  Now a total of 64 masks is
required for this action which is still regarded as rather local.  

The introduction of even the most local non-planar paths can have a
serious detrimental effect on the level of parallelism that is
possible.  It is easy to see that one can rapidly eliminate all sites
in an elementary $n \times n \times n$ cube with non-planar loops,
leading to $n^3$ masks per link direction.

\section{Summary and Conclusion}
\label{Conclusion}

We have briefly described the concept of improved actions and have
explained the implications of the non locality arising from the
improvement program for the implementation of these actions on
parallel computing architectures.  We have characterized these
implications in terms of the number of masks, which in turn determine
the minimum number of serial calculations needed to perform a Monte
Carlo updating sweep over all of the gluon links on the lattice.  We
have systematically built up a completely general algorithm using
masks that allow one to put any planar improved lattice action on a
parallel machine in an efficient way.  The generalized masking
construction are given in Sec.~\ref{maskingsupimpsu3}.

Non-planar considerations encountered in nesting specific planar
actions and actions involving non-planar loops have also been
addressed.  We hope that the methodology presented will allow one to
find an efficient parallel mask for any desired action.  We are
currently testing our algorithms on some highly improved actions and
will be reporting the results of these studies\cite{bilson-thompson}
in the near future.

\section{Acknowledgment.}

This research was supported by the Australian Research Council
and by grants of supercomputer time on the CM-5 made available through
the South Australian Centre for Parallel Computing.
AGW also acknowledges support from the U.S.\ Department of Energy
Contract No.\ DE-FG05-86ER40273 and by the Florida State University
Supercomputer Computations Research Institute which is partially
funded by the Department of Energy through Contract No.\
DE-FC05-85ER2500.

\end{document}